# Dislocation-induced thermal transport anisotropy in single-crystal group-III nitride films

Bo Sun[1,*], Georg Haunschild[2], Carlos Polanco[3], James (Zi-Jian) Ju[2], Lucas Lindsay[3], Gregor Koblmüller[2], and Yee Kan Koh[1,4,†]

[1]Department of Mechanical Engineering, National University of Singapore, Singapore 117576

[2]Walter Schottky Institute, Department of Physics, Technical University of Munich, 85748 Garching, Germany

[3]Materials Science and Technology Division, Oak Ridge National Laboratory, Oak Ridge, Tennessee 37831, USA

[4]Centre of Advanced 2D Materials, National University of Singapore, Singapore 117542

* Present Address: Division of Engineering and Applied Science, California Institute of Technology, Pasadena, California 91125, USA

† email: mpekyk@nus.edu.sg

*This manuscript has been authored by UT-Battelle, LLC under Contract No. DE-AC05-00OR22725 with the U.S. Department of Energy. The United States Government retains and the publisher, by accepting the article for publication, acknowledges that the United States Government retains a non-exclusive, paid-up, irrevocable, world-wide license to publish or reproduce the published form of this manuscript, or allow others to do so, for United States Government purposes. The Department of Energy will provide public access to these results of federally sponsored research in accordance with the DOE Public Access Plan(http://energy.gov/downloads/doe-public-access-plan).*

**Dislocations, one-dimensional lattice imperfections, are common to technologically important materials such as III-V semiconductors[1], and adversely affect heat dissipation in e.g., nitride-based high-power electronic devices[2]. For decades, conventional models[3-5] based on nonlinear elasticity theory have predicted this thermal resistance is only appreciable when heat flux is perpendicular to the dislocations. However, this dislocation-induced anisotropic thermal transport has yet to be seen experimentally[6-9]. In this study, we measure strong thermal transport anisotropy governed by highly oriented threading dislocation arrays along the cross-plane direction in micron-thick, single-crystal indium nitride (InN) films. We find that the cross-plane thermal conductivity is more than tenfold higher than the in-plane thermal conductivity at 80 K when the dislocation density is on the order of ~$3\times10^{10}$ cm$^{-2}$. This large anisotropy is not predicted by the conventional models[3,4]. With enhanced understanding of dislocation-phonon interactions, our results open new regimes for tailoring anisotropic thermal transport with line defects, and will facilitate novel methods for directed heat dissipation in thermal management of diverse device applications.**

Over the past decade, accurate experiments[10,11] and novel theoretical methods[12-16] have significantly advanced knowledge of lattice imperfections (point defects, dislocations, grain boundaries, etc.) and how these impede thermal transport in crystals and nanostructures. This in-depth understanding has facilitated better thermal management of electronic and optoelectronic devices[17], design of novel thermoelectric materials[18] and development of sophisticated technologies such as heat assisted magnetic recording[19] and phononic devices[20]. Unlike other defects, the role of dislocations in thermal resistance is still poorly understood. From the theoretical perspective, predictive first-principles calculations[12,13,16] of phonon scattering by dislocations is still nascent, partly due to the large supercells required for their description. Thus, most of the recent theoretical efforts still rely on conventional nonlinear elasticity models[3-5,8], pioneered by Klemens in the mid-1950s, to describe dislocation-phonon interactions.

According to these conventional theories, phonons are elastically scattered by dislocations via two distinct mechanisms: static scattering[3-5] and dynamic scattering[8,21]. Dynamic scattering occurs when mobile dislocations resonantly absorb an incident phonon, vibrate and re-emit a phonon through the process. To have resonant phonon-dislocation interactions, the phonon wavelengths must be comparable to the distance between two pinning points in the dislocations[7]. Phonons with such characteristically long wavelengths are important for heat conduction only at low temperatures (e.g., <10 K), and thus dynamic scattering is insignificant for heat transport at elevated temperatures[8]. Static scattering, on the other hand, can arise from the cores of the dislocations as well as from the strain field generated by these dislocations. At short range, phonons are scattered by the distortion of the lattice in the immediate vicinity of the cores. At long range, phonons are scattered by the anharmonicity related to the inhomogeneous strain fields induced by the dislocations. Klemens first treated this static scattering using perturbation theory, and found that scattering by the long-range strain fields is much stronger than scattering by the cores at temperatures lower than the Debye temperature[3]. Carruthers used a more rigorous strain field displacement in the scattering matrix, and derived a scattering cross-section ~1000 times larger than the expression by Klemens for edge-type dislocations[4]. Neither of these conventional models, however, universally and quantitatively describe the set of experimental data currently available[8,21].

One distinct prediction of the nonlinear elasticity models is that only phonons that propagate perpendicular to the dislocations are strongly scattered, due to the planar strain field generated by both edge-type and screw-type dislocations[3,5]. This phonon scattering anisotropy could give rise to large anisotropy in thermal conductivity if dislocations are highly oriented. While large anisotropy in thermal conductivity has been observed in crystals with layered structures (e.g., graphite[22], transition metal dichalcogenides[23] and black phosphorus[24]), this behavior originates mainly from the anisotropic crystal structures (e.g., anisotropy in phonon velocity[23,24]) and not from the anisotropy in scattering by extrinsic defects. The predicted anisotropy in thermal transport due to dislocation-phonon interactions has not been verified by experiment, due to the challenges of synthesizing single-crystal materials with a large density of highly-oriented dislocations and of accurately measuring their thermal conductivity, both cross-plane and in-plane. Previous measurements to study phonon-dislocation interactions were performed on bulk samples with a low density of uncontrolled, random dislocations, and thus only isotropically reduced thermal conductivities were reported[6-9].

In this report, we experimentally demonstrate strongly anisotropic thermal transport in single-crystalline semiconductors with highly oriented dislocation arrays. In particular, we exploit here the unique properties of wurtzite, i.e., (0001)-oriented group-III nitrides which form vertically well-oriented threading dislocations along the growth direction upon heteroepitaxy on lattice-mismatched substrates, exemplified here for the case of InN on GaN. Our results provide the first unambiguous evidence for dislocation-induced anisotropic thermal transport, and thus confirms this prediction by Klemens from six decades ago[3]. Furthermore, comparison of the thermal conductivity of the as-measured InN with that obtained from intrinsic *ab-initio* calculations coupled with empirical models for phonon-dislocation interactions, demonstrates that scattering strengths according to Carruthers' model are crudely comparable to the scattering strengths observed in our measurements, far better than predictions of Klemens' model. However, neither model captures the temperature dependence of our measurements accurately. We provide possible explanations for the stronger-than-expected scattering observed at lower temperatures.

Insights from this work will have technological impacts in thermal management of power electronics and directed heat spreading in electronic devices[17]. III-nitride semiconductors, especially GaN and InN, are promising materials in a wide range of emerging applications, such as power electronics[2] and light-emitting diodes (LEDs)[1]. Efficient thermal management of these high-power-density devices requires detailed understanding of phonon scattering by dislocations, commonly found in III-nitrides due to limitations of current growth techniques[25]. Also, we present a new route to anisotropically direct heat flow via dislocations, which could be used to guide and spread heat in electronic devices[17]. Different from previous anisotropic heat transport that relies on material crystal structures, the anisotropic heat transport presented here originates from crystal defects, which may be manipulated and implanted independent of crystal geometry.

We synthesized ~1.5-2 μm-thick, heteroepitaxial wurtzite InN films on c-plane, (0001)-oriented GaN/sapphire substrates (Figure 1a) using plasma-assisted molecular beam epitaxy (PAMBE)[26,27]. We determine the density of threading dislocations in our InN films from X-ray rocking curves (ω scans) and cross-sectional transmission electron microcopy (TEM) under the two-beam condition, see Figures 1b – 1e and the Supplementary. With the TEM images taken with different diffraction conditions, edge-type (Figures 1b and 1d) and screw-type (Figures 1c and 1e) threading dislocations can also be selectively imaged and counted[26]. The density of edge-type dislocations is found to be more than 20 times larger than that of screw-type dislocations, in agreement with recent studies of heteroepitaxial InN films grown on GaN/sapphire substrates[26,28]. This suggests that the prominent role of dislocation-phonon interactions and effects on thermal conductivity seen in our materials is mainly due to edge-type dislocations.

We measure the in-plane thermal conductivity $\Lambda_\parallel$ (perpendicular to dislocations) and the cross-plane thermal conductivity $\Lambda_\perp$ (along the dislocations) of the InN films using time-domain thermoreflectance (TDTR)[29,30]. Details of the implementation, analysis and uncertainty estimation of the TDTR measurements are provided in Methods and Supplementary Information. We note that our approach includes a correction that accounts for artifacts due to the leaked pump beam from rough sample surfaces[30].

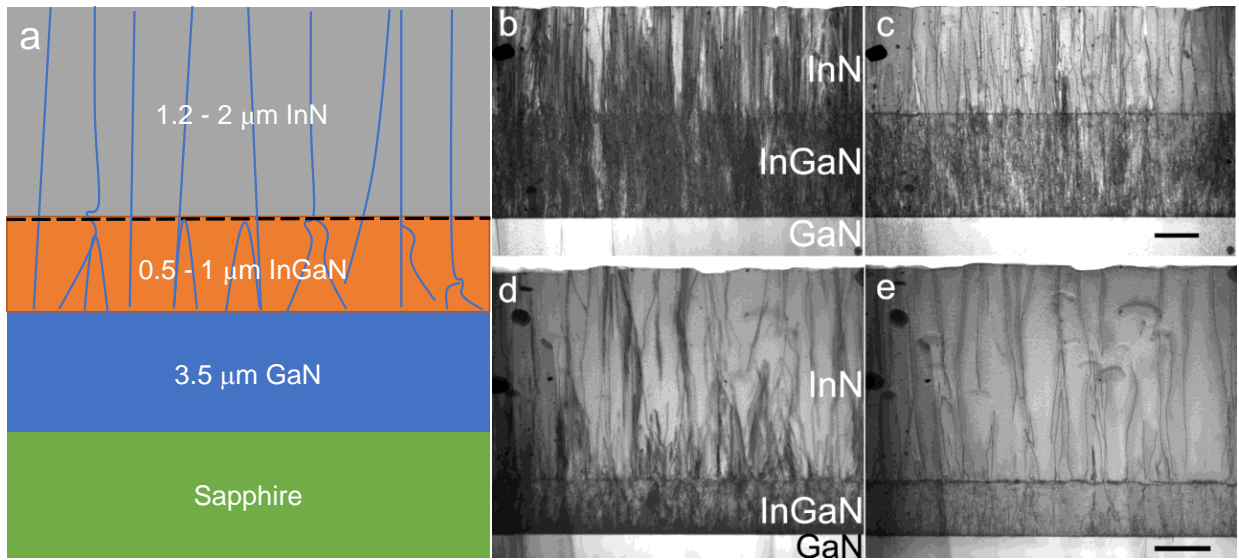

**Figure 1 | a)** Schematic illustration of our sample structure. **b - e)** Cross-sectional TEM images of the InN films with threading dislocation densities of $1.1 \times 10^{10}$ cm$^{-2}$ (b & c) and $2.9 \times 10^{10}$ cm$^{-2}$ (d & e). Edge-type (b & d) and screw-type (c & e) threading dislocations are imaged at diffraction conditions of $g = 1\bar{1}00$ and $g = 0002$, respectively. The scale bars are 500 nm.

Figure 2a gives the measured in-plane ($\Lambda_\parallel$) and cross-plane ($\Lambda_\perp$) thermal conductivities of our InN films as a function of temperature (T). Contrary to the small anisotropy induced by other extrinsic defects (e.g., boundary scattering in superlattices[31]), we observe a large dislocation-induced anisotropy in the low-T thermal conductivity of our InN films with oriented threading dislocations, which is very sensitive to T, see Fig. 2b. The intrinsic anisotropy of InN due to its wurtzite structure is small and independent of T, with an anisotropy ratio[13] of only 1.17 (see our first-principles calculations in Figure 2b). We contrast the anisotropic thermal conductivity of our InN films with previously reported, structurally-induced anisotropic thermal conductivities of other materials in Figure 2b[22-24,31-33]. The anisotropy we observe is fundamentally different in three important aspects. First, previous reports demonstrated structurally-induced anisotropies that are not easily manipulated, while here we demonstrate defect-induced thermal conductivity anisotropy that can be engineered. Second, all prior reported measurements have suppressed thermal conductivity in the through-plane crystallographic orientation, while the measured thermal conductivity of our InN films is subdued along the basal plane, see Figures 2a and 2b (open *vs.* solid symbols). In the cross-plane direction, $\Lambda_\perp$ of our InN films approaches the intrinsic thermal conductivity of InN predicted by our first-principles calculations, with a $T^{-1}$ temperature dependence. Meanwhile, in the in-plane direction, $\Lambda_\parallel$ of our InN films is strongly suppressed at low T. Third, as a result of the unique T dependence of $\Lambda_\parallel$ in our InN films, the anisotropy ratios of our measurements exhibit a strong T dependence, see Figure 2b. This T dependence is starkly different from prior reported thermal conductivity anisotropies, which are largely T independent[22,24]. At room temperature, $\Lambda_\parallel$ and $\Lambda_\perp$ of all our InN films are similar within experimental uncertainty. However, at low T, we observe large anisotropy for all our InN films, the extreme case being $\Lambda_\perp/\Lambda_\parallel = 14$ at 80 K for the InN film with a dislocation density of $2.9\times10^{10}$ cm$^{-2}$.

The differences between our work and previously reported measurements demonstrate that thermal conductivity anisotropies as observed by the present InN films have a different origin. Prior reported thermal conductivity anisotropies originate from the anisotropies in crystal structure[23,24]. Substantially larger thermal conductivities along the basal plane were reported for crystals with highly anisotropic phonon dispersions (and thus anisotropic phonon velocity), such as black phosphorus[24] and graphite[22]. Moreover, anisotropic crystal structures could also result in anisotropic scattering of phonons. For example, we previously demonstrated that anisotropy in the thermal conductivity of black phosphorus in the basal planes is mainly due to anisotropy in phonon dispersion, while the low through-plane thermal conductivity partially originates from enhanced Umklapp scattering across the basal planes[24]. For these structure-induced anisotropies, phonons are predominantly scattered in all crystallographic directions by the same phonon-phonon scattering processes with the same T dependence[13,24]. As a result, the anisotropy ratios of these crystals are largely T independent, see Figure 2b.

On the contrary, the unique T dependent anisotropy we observe in our measurements indicates that phonons are predominantly scattered by different mechanisms in the in-plane and cross-plane directions of InN films with highly oriented threading dislocations. In the cross-plane direction, we observe that $\Lambda_\perp$ is proportional to T$^{-1}$. This T dependence suggests that in the cross-plane direction phonons are predominantly scattered by three-phonon Umklapp processes[13,24]. On the other hand, in the in-plane direction, we attribute the strong suppression of $\Lambda_\parallel$ at low T to the anisotropic scattering by dislocations. The suppression of heat conduction only in the in-plane direction cannot be explained by other extrinsic scattering mechanisms, such as point defects, interfaces and grain boundaries, as these scattering mechanisms inevitably impede heat flow in both in-plane and cross-plane directions[32,33]. While boundary scattering could result in thermal conductivity anisotropies (see the anisotropy of GaAs/AlAs superlattices[31] and Si thin films[32,33] in Figure 2b) the anisotropy is relatively small and thermal transport is more strongly suppressed in the cross-plane direction[31-33]. Hence, our results unambiguously demonstrate dislocation-induced thermal conductivity anisotropy as predicted by Klemens[3].

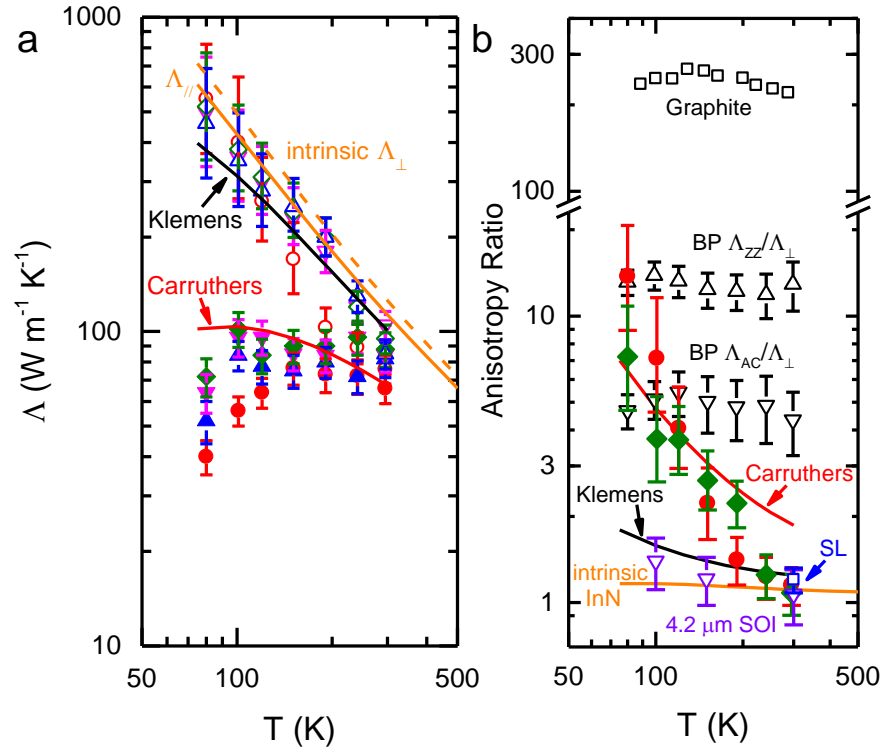

**Fig. 2 | Temperature dependent thermal conductivity of InN films with oriented threading dislocations. a)** In-plane $\Lambda_\parallel$ (solid symbols) and cross-plane $\Lambda_\perp$ (open symbols) thermal conductivity of InN films with dislocation densities of $1.1\times10^{10}$ cm$^{-2}$ (olive diamonds), $1.5\times10^{10}$ cm$^{-2}$ (magenta down triangles), $2.5\times10^{10}$ cm$^{-2}$ (blue up triangles) and $2.9\times10^{10}$ cm$^{-2}$ (red circles), respectively. The intrinsic thermal conductivities $\Lambda_\parallel$ (dashed orange curve) and $\Lambda_\perp$ (solid orange curve) of InN are calculated from first principles. The in-plane thermal conductivity including phonon-dislocation scattering from the Carruthers' model[4] (red curve) and the Klemens' model[3] (black curve), for a dislocation density of $2.9\times10^{10}$ cm$^{-2}$, are included for comparison. There are no adjustable parameters in the empirical calculations. **b)** Temperature dependent anisotropy ratios of InN films with a dislocation density of $1.1\times10^{10}$ cm$^{-2}$ (olive diamonds) and $2.9\times10^{10}$ cm$^{-2}$ (red circles), compared to the anisotropy

ratios of graphite (black squares, ref. 22), black phosphorus (black triangles, ref. 24, the in-plane thermal conductivities in zigzag and armchair directions are labeled as $\Lambda_{ZZ}$ and $\Lambda_{AC}$ respectively), GaAs/AlAs superlattice (blue square, ref. [31]) and 4.2-µm-thick SOI (purple triangles, ref. [32,33]). The solid symbols represent $\Lambda_\perp/\Lambda_\parallel$, while the open symbols represent $\Lambda_\parallel/\Lambda_\perp$. (Here, for crystals, $\Lambda_\parallel$ represents the thermal conductivity along the basal planes and $\Lambda_\perp$ represents the thermal conductivity through the basal planes.) The anisotropy ratios of the intrinsic thermal conductivity of InN from our first-principles calculations (orange curve), and with a dislocation density of $2.9\times10^{10}$ cm$^{-2}$ calculated using the Klemens' model (black curve) and the Carruthers' model (red curve), are also presented.

To obtain a clearer picture of the effect of threading dislocations on thermal transport, we show both $\Lambda_\parallel$ and $\Lambda_\perp$ as a function of dislocation density in Figure 3. At room temperature, $\Lambda_\parallel$ and $\Lambda_\perp$ are similar for all samples despite the variation in dislocation density. We note that the overall very high dislocation density of our samples (~$10^{10}$ cm$^{-2}$) is insufficient to diminish the intrinsic thermal conductivity of InN appreciably, indicating that phonon-dislocation scattering is weaker than the intrinsic phonon-phonon scattering with this dislocation density. This observation agrees with a recent molecular dynamics calculation[34], which predicts that an apparent thermal resistance due to dislocations at room temperature can only be observed in GaN when the dislocation density is at least $10^{12}$ cm$^{-2}$. For $\Lambda_\perp$, we do not observe any noticeable change as a function of the dislocation density even at 80 K, suggesting that phonons are not strongly scattered along the dislocation lines. For $\Lambda_\parallel$, we observe a much stronger dependence on dislocation density at 80 K, consistent with our argument that the suppression of the in-plane thermal conductivity at lower temperatures is due to oriented threading dislocations.

Our measurements of anisotropic thermal conductivity provide critical data to understand anisotropic dislocation-phonon interactions. To further examine these interactions and their role in determining anisotropic thermal transport, we calculate the thermal conductivity of the InN films with oriented threading dislocations from a full solution of the Peierls-Boltzmann transport equation combining *ab-initio* three-phonon and isotope scattering rates with different empirical models for phonon-dislocation scattering rates, see Methods section for details. We only consider static scattering by dislocation cores and long-range strain fields using Klemens' and Carruthers' models[3,4], as dynamic scattering by dislocations is significant only at low T. Our calculations indicate that the commonly used Klemens' model[3] underestimates the strength of phonon-dislocation scattering, resulting in thermal conductivity values over an order of magnitude larger than experimental data at low T, see Figure 2a. This deficiency, also found in LiF samples[6], motivated Carruthers to use a logarithmic strain field displacement, which increases the scattering strength[4]. Figure 2a shows the prediction of thermal conductivity when the Carruthers model is used. The *order-of-magnitude* similarity between calculated values and experimental data around room temperature is impressive given the many simplifications of the model and absence of fitting parameters. Nevertheless, Carruthers' model under-predicts the scattering strength by dislocations especially at low T, and thus does not capture the T behavior of thermal conductivity for different dislocation densities (Fig 2a).

We consider a few possible reasons for the failure of Carruthers' model to capture the T dependence of the in-plane thermal conductivity of the InN films. One possible explanation is omission of scattering by the dislocation cores in Carruthers' model. However, we find that scattering by dislocation cores using the Klemens model is negligible, even at room temperature where these are predicted to be more important. (See the calculations with and without scattering by dislocation cores in figure S3-2 in the supplementary information.)

Another possible explanation of the weaker scattering at low T predicted by the empirical models is neglecting the overlapping of dislocation strain fields, as the models only treat phonon scatterings from single dislocations, likely valid at low dislocation concentrations. The collective effects of overlapping strain fields at higher concentrations may scatter long-wavelength phonons preferentially. In our InN films, the dislocation densities of our threading dislocation arrays are large, with an average inter-dislocation spacing of only 60 – 100 nm. At low T, the wavelength of phonons in InN that carry the majority of heat could be comparable to the inter-dislocation spacing, and thus the assumption of isolated dislocations is violated. To evaluate this possibility, we plot the accumulated thermal conductivity as a function of phonon wavelength in figure S3-1. We find that the wavelengths of heat-carrying phonons are mostly <5 nm even at 80 K, not sufficiently large to be coupled with the periodic strain fields in our samples.

Lastly, the empirical models might under-estimate the scattering rate due to negligence of change of interatomic forces in the vicinity of dislocation cores. A recent *ab-initio* calculation of phonon scattering by the cores of dislocations, without considering long-range strain field effects, showed larger scattering rates than those predicted by analytical models for Si[16]. These larger scattering rates were attributed to changes in the interatomic force constants in the neighborhood of the dislocation, which are not properly treated by first order perturbation approaches[14-16,35] in Klemens' and Carruthers' models. Similar *ab-initio* methods could provide insights here; however, these are beyond our current computational resources due to the large supercells required to model dislocations with long-range strain field.

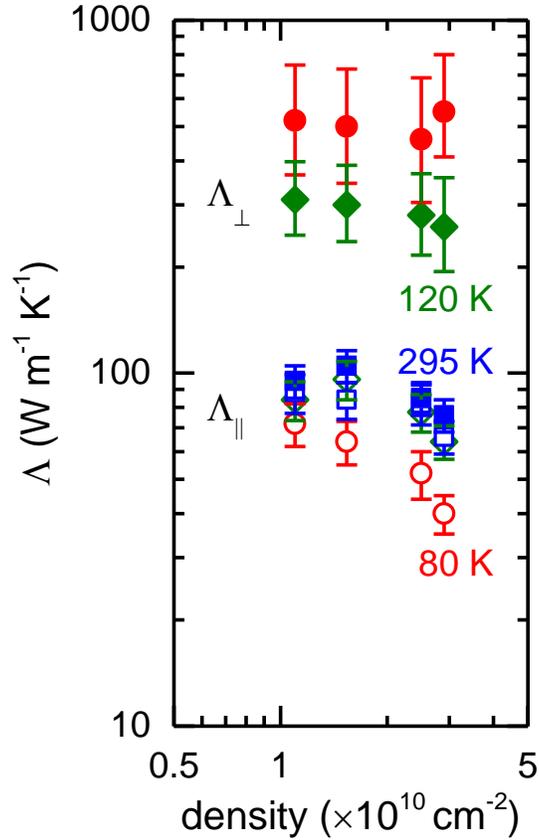

**Fig. 3 | In-plane (open symbols) and cross-plane (solid symbols) thermal conductivity of InN films as a function of dislocation density at 80 K (red), 120 K (olive) and 295 K (blue).**

In conclusion, we observe unambiguous evidence for dislocation-mediated anisotropic thermal transport in single-crystal InN films with highly oriented threading dislocations. Our results validate predictions of this, though show stronger phonon-dislocation interactions than predictions of conventional models especially at low temperatures. Due to strong dislocation-phonon interactions, we observe an anisotropy ratio of up to 14 in the thermal conductivity of InN films at 80 K. Our results adds an important degree of freedom for the thermal management of nitride-based electronic and optoelectronic devices, and novel designs for directed heat dissipation for electronic devices.

## Methods

### InN film synthesis with highly oriented threading dislocations.

Plasma-assisted molecular beam epitaxy (PAMBE) was used to grow In-polar InN thin films on c-plane, (0001)-oriented GaN/sapphire substrates employing an $In_{0.8}Ga_{0.2}N$ buffer layer. PAMBE growth of InN films on c-plane GaN is well known to produce single-crystalline films[36,37], however, with a high density of threading dislocation defects due to the large lattice mismatch[26,27,28]. Both the InGaN buffer layer and subsequent InN film were grown under metal-rich, i.e., In-rich growth conditions to produce single-crystalline films with smooth surface morphology and high-quality, electronic-grade properties[26,27,36,38]. Due to the metal-rich conditions, macroscopic In-droplets typically accumulate at the growth surface[38], which were subsequently removed in buffered hydrochloric (HCl) acid to produce specular, atomically smooth surfaces for TDTR measurements.

In our sample structure, we implemented a highly In-rich $In_xGa_{1-x}N$ buffer layer (x = 0.8) (see Figure 1a) that serves two purposes. First, the low thermal conductivity of this buffer layer forces heat to flow in the in-plane direction within the InN films. As a result, heat dissipation in our measurements is sensitive to the in-plane thermal conductivity of InN even at 80 K, and in-plane thermal conductivities can be measured accurately. Second, the high In-content $In_{0.8}Ga_{0.2}N$ buffer layer was implemented to reduce the lattice mismatch between the epitaxial InN film and the GaN substrate, such that the largest portion of mismatch strain is relieved within the buffer layer. Thereby, non-vertically oriented threading dislocations, which also form during heteroepitaxial nucleation of group-III nitride films near the substrate interface[26,39], are readily annihilated within the $In_{0.8}Ga_{0.2}N$ buffer layer – a process that is driven by the merging of two inclined threading dislocations with opposite Burgers vector[39]. As a result, the largest fraction of threading dislocations propagating in the InN film are relatively well-oriented along the vertical *c*-axis (i.e., (0001) orientation), which we later refer to as the cross-plane direction. As shown in Figures 1b – 1e, inclined dislocations are also formed, however, they are predominantly located in the $In_{0.8}Ga_{0.2}N$ buffer layer where they fuse and annihilate[39].

In order to realize variation in the dominant edge-type threading dislocation density we varied the growth temperature for the InN films between 450°C and 500°C, in accordance to Ref. 36. Hereby, higher growth temperature yields lower dislocation density compared to low growth temperatures. In total, we explored four samples with different threading dislocation densities in the InN films, ranging from $1.1\times10^{10}$ cm$^{-2}$ to $2.9\times10^{10}$ cm$^{-2}$.

**Time-domain thermoreflectance (TDTR).**

We coated a thin Al layer (~100 nm) on the samples for TDTR measurements. Laser pulses from an ultrafast laser are split into a pump and a probe beam. The pump beam heats the samples periodically, creating a temperature oscillation. The probe beam monitors the temperature oscillation at the sample surfaces through thermoreflectance (i.e., change of reflectance with temperature). Since the induced temperature oscillation depends on the thermal properties of each sample, the thermal conductivity can be extracted. In this study, we used a $1/e^2$ laser radius of 5 μm and a modulation frequency of 0.5 MHz for in-plane thermal measurements, while we used 28 μm and 10 MHz for cross-plane thermal measurements. To measure cross-plane thermal conductivity at 80 K, 100 K and 120 K, we used a modulation frequency of 22.2 MHz for improved sensitivity. The uncertainty of our measurements of in-plane and cross-plane thermal conductivity is estimated as 12 % and 10 % at room temperature, and 13 % and 40 % at 80 K, respectively. Details can be found in the supplementary information.

**Density functional theory calculations.**

The thermal conductivity of in-plane and cross-plane InN is calculated solving self-consistently the Peierls-Boltzmann transport (PBT) equation[40-42]. Three-phonon and phonon-isotope scattering rates are considered within first order perturbation theory as described in previous numerical works[12,43-47]. These rates are combined with empirical phonon-dislocation scattering rates derived by Klemens[3] and Carruthers[4] using Matheissen's rule, and then solving the PBT equation. Further details of the implementation of this methodology and description of the empirical phonon-dislocation scattering rates can be found in the supplementary information.

**See supplementary information for more information.**

**Acknowledgements**

We thank Rui Wang and Bin Huang at NUS for help on thermal evaporation of Al films. We would like to thank Professor Mingda Li at MIT for explanation of his papers.  This work was supported by NUS Start-up Grant, the Singapore Ministry of Education Academic Research Fund Tier 2 under Award No. MOE2013-T2-2-147 and Singapore Ministry of Education Academic Research Fund Tier 1 FRC Project FY2016. C.P. and L.L. acknowledge support from the U. S. Department of Energy, Office of Science, Office of Basic Energy Sciences, Materials Sciences and Engineering Division and computational resources from the National Energy Research Scientific Computing Center (NERSC), a DOE Office of Science User Facility supported by the Office of Science of the U. S. Department of Energy under Contract No. DE-AC02-05CH11231. G.K. gratefully acknowledges support from the excellence program Nanosystems Initiative Munich (NIM) funded by the German Research Foundation (DFG).

**Author contributions**

G.K. and Y.K.K. initialized the idea, B.S. and Y.K.K. designed the experiments, B.S. performed the TDTR measurements and analyzed the data, G.H., J.Z.J. and G.K. prepared and characterized

**Additional information**

Supplementary information is available.

**Competing financial interests**

The authors declare no competing financial interests.

# Supporting Information

# Dislocation-induced thermal transport anisotropy in single-crystal group-III nitride films

Bo Sun[1, *], Georg Haunschild[2], Carlos Polanco[3], James (Zi-Jian) Ju[2], Lucas Lindsay[3], Gregor Koblmüller[2], and Yee Kan Koh[1, 4, †]

[1]Department of Mechanical Engineering, National University of Singapore, Singapore 117576

[2]Walter Schottky Institut, Physik Department, TU München, 85748 Garching, Germany

[3]Materials Science and Technology Division, Oak Ridge National Laboratory, Oak Ridge, Tennessee 37831, USA

[4]Centre of Advanced 2D Materials, National University of Singapore, Singapore 117542

* Present Address: Division of Engineering and Applied Science, California Institute of Technology, Pasadena, California 91125, USA

† email: mpekyk@nus.edu.sg

## 1 TDTR measurements of thermal conductivity

### 1-1 Sensitivity to in-plane thermal conductivity of InN films with and without InGaN barrier layer

Generally, a small spot size and a low modulation frequency are preferred for measurements of in-plane thermal conductivity using TDTR, as this configuration facilitates in-plane heat diffusion. However, this is only true when the induced thermal wave does not propagate deep into the substrate during the TDTR measurement cycle. In our case, GaN is a good thermal conductor and the sensitivity of TDTR measurements to the in-plane thermal conductivity of InN films would be small if there is no barrier layer. This is especially true for measurements at lower temperatures when the thermal conductivity of GaN is large. To overcome this challenge, an InGaN layer was introduced to enhance the sensitivity to in-plane thermal measurements. Here we calculated the sensitivity of the in-plane thermal conductivity of InN films with and without an InGaN buffer layer with the parameters for in-plane measurements [1]. We define measurement sensitivity to in-plane thermal conductivity $\Lambda_\parallel$ as:

$$S_{\Lambda_\parallel} = \frac{\partial \ln R}{\partial \ln \Lambda_\parallel}$$

where R is the absolute value of the ratio of in-phase and out-of-phase signal in TDTR measurements. A high value of $S_{\Lambda_\parallel}$ means that the TDTR measurements are sensitive to the in-plane thermal conductivity and are reasonably accurate. We plot $S_{\Lambda_\parallel}$ as a function of thickness of InGaN buffer layer ($h_{InGaN}$) in Figure S1-1. We can see from the figure that without an InGaN barrier layer, the sensitivity to $\Lambda_\parallel$ is rather low, down to ~0.07 at 80 K. Such low sensitivity will lead to large error bars in the thermal conductivity measurements. With 500-nm thick InGaN, however, sensitivity to $\Lambda_\parallel$ increases to ~0.35 at 80 K. With such sensitivity, an experimental error of <15 % is achieved in the measurements of $\Lambda_\parallel$.

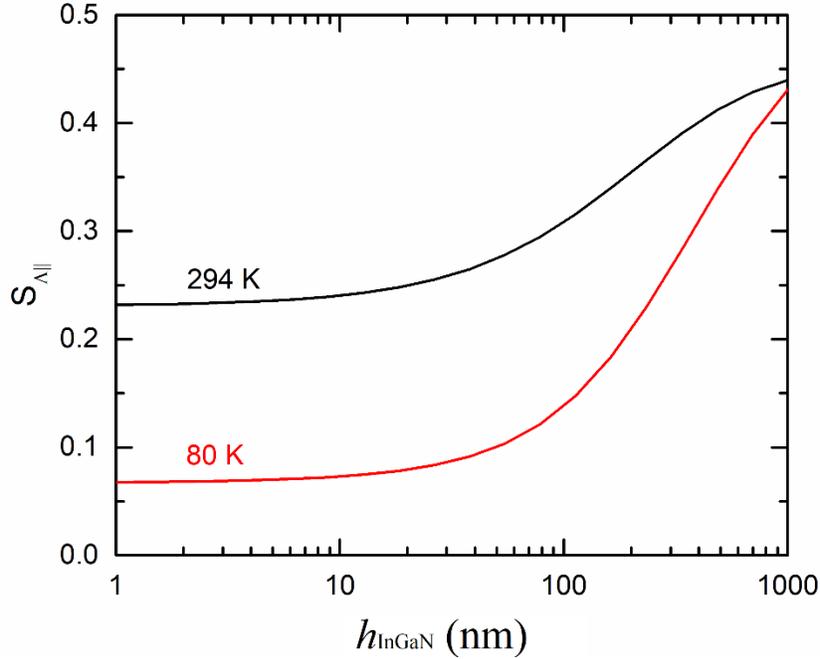

Fig. S1-1 Measurement sensitivity to in-plane thermal conductivity of InN at room temperature (black) and at 80 K (red) as a function of thickness of the InGaN layer. The modulation frequency is 0.5 MHz and the laser spot size ($1/e^2$ radius) is 5 μm for in-plane measurements.

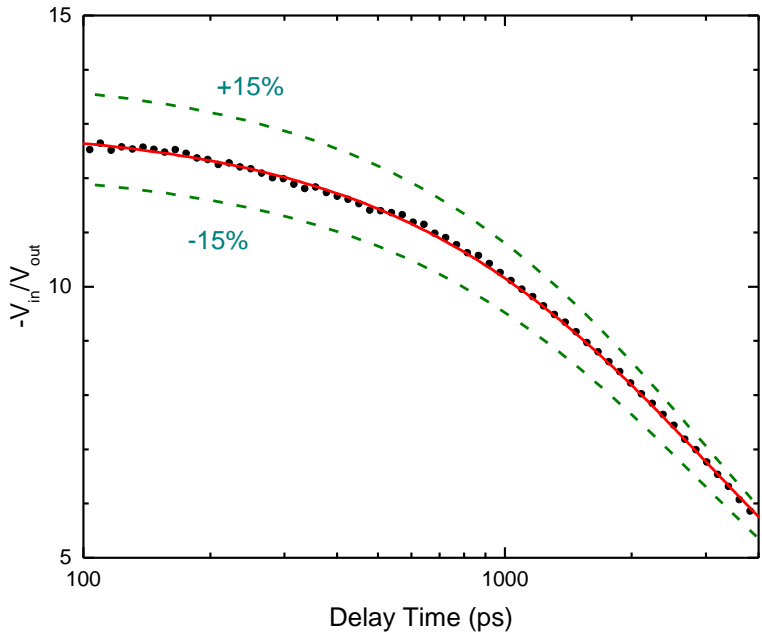

Fig. S1-2 Measurement of the in-plane thermal conductivity of an InN film with dislocation density of $1.1 \times 10^{10}$ cm$^{-2}$ at 80 K. The solid curve is the best fit using thermal modeling with $\Lambda_{||} = 72$ W m$^{-1}$ K$^{-1}$. The two dashed curves are calculated using thermal modeling with $\Lambda_{||} = 72 \pm 15\%$ W m$^{-1}$ K$^{-1}$.

## 1-2 TDTR measurement of cross-plane thermal conductivity

The cross-plane measurements of thermal conductivity are only challenging when the thermal resistivity of the InN film is small. In this case, the measured cross-plane thermal conductivity has a large uncertainty especially at low temperatures. To improve the accuracy of our measurements, we increased the modulation frequency to 22.2 MHz for TDTR measurements of cross-plane thermal conductivity at 80 K – 120 K, to reduce this uncertainty. The laser spot size ($1/e^2$ radius) was 28 μm. In Figure S1-3 we show an example of large uncertainty at 80 K for cross-plane thermal conductivity measurements. The large uncertainty is reflected in the error bars in Figs. 2 and 3 of the main manuscript.

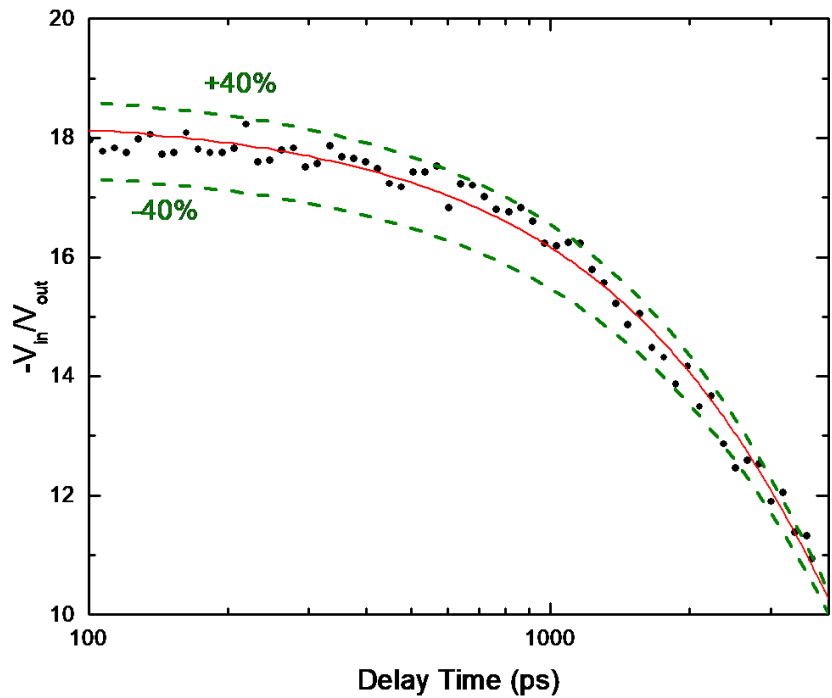

**Fig. S1-3: TDTR measurement of the cross-plane thermal conductivity of an InN thin film with dislocation density of $1.1 \times 10^{10}$ cm$^{-2}$ at 80 K. Modulation frequency of the pump beam is 22.2 MHz and the laser spot size ($1/e^2$ radius) is 28 μm.**

## 2 Characterization of InN films
### 2-1 X-ray diffraction (XRD) and X-ray rocking curve measurements

Epitaxial single-crystalline growth of InN/InGaN on c-plane (0001) GaN template was confirmed by XRD 2θ-ω scans, see Figure S2-1. From Fig. S2-1, we can see all respective diffraction peaks associated with the individual layers, i.e., indexed as InN (0002), InGaN (0002) and GaN (0002), indicating that all layers are [0001]-oriented, wurtzite phase. The InN and InGaN diffraction peaks are well separated from the GaN substrate peak, confirming that the mismatch strain is fully relieved. From the 2θ peak position of the InGaN buffer layer, we further confirmed its alloy composition yielding an In-content of $x(In) = 0.78$, close to the nominally expected value.

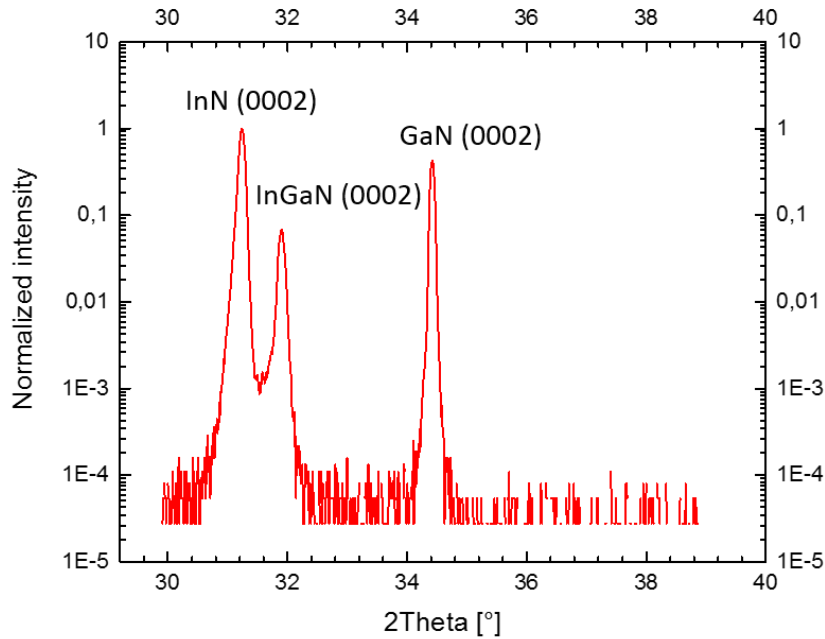

**Fig. S2-1: XRD 2θ-ω scan of InN sample with dislocation density of $1.5 \times 10^{10}$ cm$^{-2}$.**

X-ray rocking curves (ω-scans) are used to estimate the density of threading dislocations. We performed X-ray rocking curve (XRC) measurements of the off-axis ($20\bar{2}1$) reflection of InN in skew symmetric geometry under open detector, which is most sensitive to the dominant mixed edge-type threading dislocations (TDs) [2]. An exemplary XRC scan is illustrated in Figure S2-2 for the InN film with the lowest investigated dislocation density ($1.1\times10^{10}$ cm$^{-2}$), giving a peak broadening (full-width at half-maximum, FWHM) of 0.46°. Note, that when measured in skew symmetric geometry, off-axis planes are disrupted by mixed edge-type threading dislocations in (0001)-oriented crystals and the FWHM of the skew symmetric rocking curve (20$\bar{2}$1) is broadened by the presence of these dislocations.

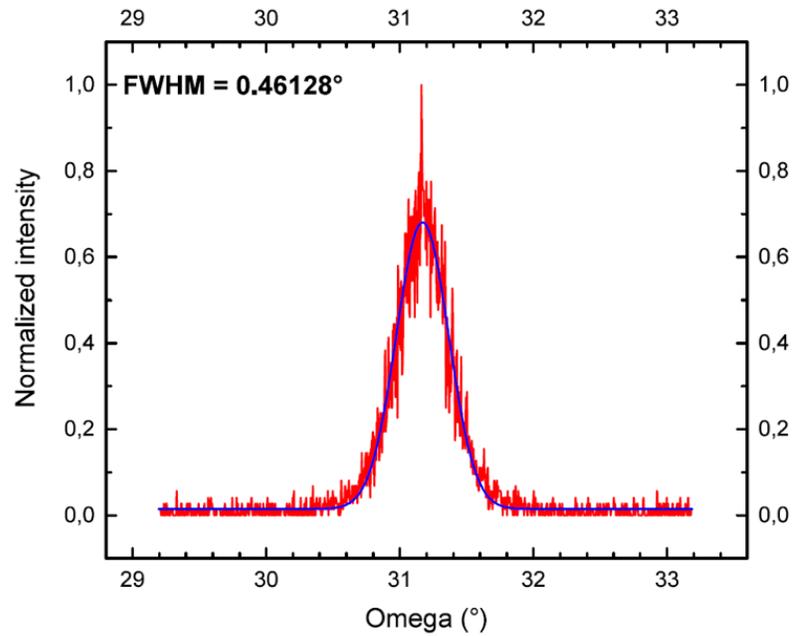

**Fig. S2-2 X-ray rocking curve (ω-scan) of off-axis (20$\bar{2}$1) reflection of the InN film with dislocation density of $1.1 \times 10^{10}$ cm$^{-2}$.**

Using the models developed by Srikant et al. [3], we then calculated the threading dislocation densities of our InN films based on the FWHM data of the off-axis (20$\bar{2}$1) XRC, which is oriented 75° from the (0001) film plane and thereby captures all threading dislocations (TD) with edge-component. The following equation is used to estimate the majority edge-type TD densities [2]:

$$\rho_e = \Gamma_z^2/1.88a^2$$

where *a* is the corresponding Burgers vector (i.e., in-plane lattice constant a = 3.538 Å for InN) for edge-type TDs $\Gamma_z$ is the theoretical peak broadening at 90° off-axis skew geometry which is approximated by the FWHM of the (20$\bar{2}$1) XRC [2]. We summarize the FWHM and the derived TD density (calibrated by TEM) in Table S2-1. Note that the TDs estimated from X-ray rocking curves include both edge-type and mixed-type TDs, as the XRC measurement of off-axis (20$\bar{2}$1) reflection of InN are sensitive to both types of TDs.

### 2-2 Transmission electron microscopy (TEM)

We verify the density of threading dislocations (TDs) of two InN films by cross-sectional transmission electron microcopy (TEM) under the two-beam condition (measured in a 200 keV-TEM by Evans Analytical Group). For beam conditions with zone axis Z = [11$\bar{2}$0] and g = [1$\bar{1}$00], edge-type TDs become visible while other types are suppressed. In contrast, under beam conditions of Z = [11$\bar{2}$0] and g = [0002], the TEM images are only sensitive to screw-type TDs [4]. The corresponding cross-sectional TEM images of two InN films are shown in Figures 1b – 1e. We further determine the total TD density by counting all dislocations from cross-sectional TEM images recorded under four-beam condition, as shown in Figure S2-3, which has an uncertainty of ~10% in the limit of such high TD densities.

We summarize the density of edge TDs, screw TDs and mixed TDs derived from the TEM images in Table S2-1. For both samples, we find that the edge-type and the mixed TDs are predominant, while screw-type dislocations only make up <2.5% of the total TDs.

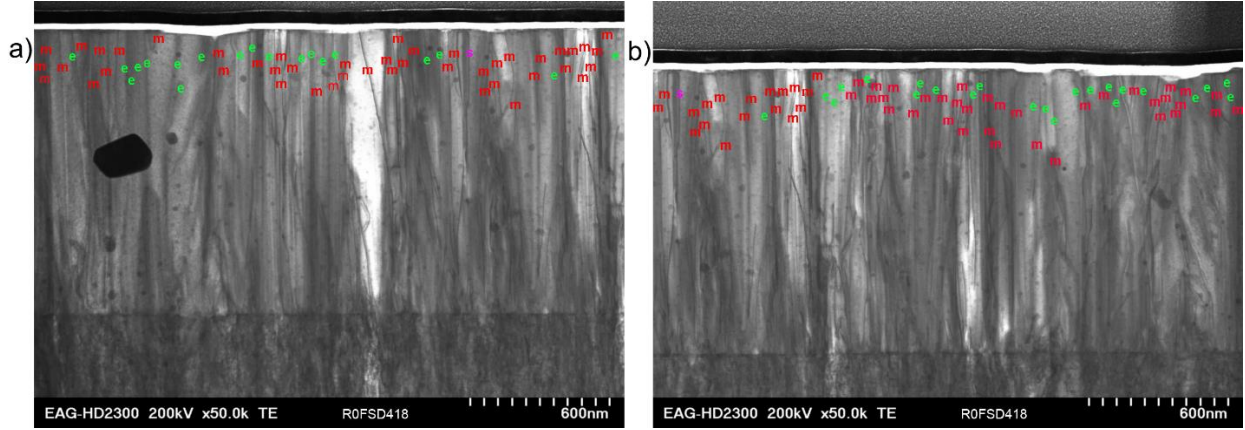

**Fig. S2-3: TEM images of a sample with a dislocation density of ~3×10$^{10}$ cm$^{-2}$ in the InN film recorded at two different regions, i.e., a) area 1 and b) area 2. The two areas are randomly chosen. Edge-type, screw-type and mixed-type dislocations are marked with green e, pink s, and red m, respectively. The dislocation type was determined using different diffraction conditions, as shown in Fig.1 in the main text.**

**Table S2-1: TD density from X-ray rocking curves and cross-sectional TEM images**

|        | X-ray rocking curves | | TEM images | | | |
|--------|---------------------|---|------------|---|---|---|
| Sample | FWHM$_{(20\bar{2}1)}$ (°) | Edge + Mixed TD density (×10$^{10}$ cm$^{-2}$) | Total TD density (×10$^{10}$ cm$^{-2}$) | Edge TD density (×10$^{10}$ cm$^{-2}$) | Screw TD density (×10$^{10}$ cm$^{-2}$) | Mixed TD density (×10$^{10}$ cm$^{-2}$) |
| 1 | 0.4613 | 1.1 | 1.1 | 0.65 | 0.03 | 0.45 |
| 2 | 0.5244 | 1.54 | - | - | - | - |
| 3 | 0.7009 | 2.53 | - | - | - | - |
| 4 |  | 2.84 | 2.9 | 0.84 | 0.04 | 2.00 |

## 3    Thermal conductivity calculations

*Intrinsic conductivity of InN (theory)*:  In-plane and cross-plane intrinsic lattice thermal conductivities ($\Lambda_{in}$ and $\Lambda_{cross}$) are given by:

$$\Lambda_\alpha = \sum_{\vec{q}j} C_{\vec{q}j} v_{\vec{q}j\alpha}^2 \tau_{\vec{q}j\alpha} \qquad (1)$$

where $C_{\vec{q}j}$ is the volume normalized mode heat capacity for phonon with wavevector $\vec{q}$ in branch $j$, $v_{\vec{q}j\alpha}$ is phonon velocity in the $\alpha$th direction and $\tau_{\vec{q}j\alpha}$ is the phonon transport lifetime in the $\alpha$th direction coinciding with the temperature gradient.  Here, the cross-plane conductivity corresponds with the z-direction coincident with the c-axis of the wurtzite structure.  The transport lifetimes are determined by full solution of the Peierls-Boltzmann transport equation[5-7] with initial inputs: three-phonon and phonon-isotope relaxation rates from quantum perturbation theory combined with empirical phonon-dislocation scattering from Mathiesen's rule: $1/\tau_{\vec{q}j} = 1/\tau_{\vec{q}j}^{3-ph} + 1/\tau_{\vec{q}j}^{ph-iso} + 1/\tau_{\vec{q}j}^{ph-disloc}$. The description of $1/\tau_{\vec{q}j}^{3-ph}$ [6, 7], $1/\tau_{\vec{q}j}^{ph-iso}$ [8] and solution of the Peierls-Boltzmann equation[9, 10] as coupled with density functional theory (DFT) methods has been described in numerous previous works[10-14], thus only details relevant to the calculation of $\Lambda$ for InN are given here.

The input parameters for the first principles $\Lambda$ and empirical phonon-dislocation scattering mechanisms are determined from DFT calculations as implemented by the plane-wave-based Quantum Espresso open source software [15].  The projector augmented wave pseudopotentials within PBE-SOL methods were used with cut-off energy of 60 Ry, electronic convergence threshold of $10^{-10}$ Ry and a Monkhorst-Pack [16] grid 8x8x6 shifted from the origin for initial structural calculations.  With these settings we obtain lattice parameters: $a$=3.538 Å, $c$=5.717 Å and $u$=0.379 (internal degree of freedom for wurtzite structure), in reasonable agreement with experiment $a_{exp}$=3.538 Å, $c_{exp}$=5.703 Å and $u_{exp}$=0.375 [17]. Elastic constants for the phonon-dislocation scattering formulas were found using the ElaStic package [18] with a Lagrangian strain 0.1.

Harmonic interatomic force constants (IFCs), Born effective charges and the high frequency dielectric tensor for determining phonon frequencies were determined via density functional perturbation theory [19] with a *k*-mesh of 6×6×4. The anharmonic IFCs for building three-phonon matrix elements were calculated to 5[th] nearest neighbors of the unit cell atoms with Γ-point-only self-consistent field calculations with perturbations (0.05 Å) in 108 atom supercells with 100 Ry energy cut-off. Point group symmetries and translational invariance were simultaneously enforced on these via a $\chi^2$ minimization algorithm [13].

*Phonon-dislocation scattering:* To model the thermal resistance from dislocations in the InN thin films we employed: (1) Klemens' (static)[20] and (2) Carruthers (static) [21]. In Klemens' theory, phonon-dislocation scattering is described by scattering from a dislocation core and from the surrounded strained elastic field and is approximated as [20, 22, 23]:

$$1/\tau_{\vec{q}j}^{ph-core} = N_d V^{4/3} \omega_{\vec{q}j}^3 / v_s^2 \qquad (2)$$

$$1/\tau_{\vec{q}j}^{ph-strain} = (\frac{2^{3/2}}{3^{7/2}}\gamma^2 \omega_{\vec{q}j})\{(N_{ds}+N_{dm})b_s^2 + (N_{de}+N_{dm})b_e^2[\frac{1}{2}+\frac{1}{24}(\frac{1-2\nu}{1-\nu})^2(1+\sqrt{2}(\frac{v_L}{v_T})^2)^2]\} \qquad (3)$$

these terms are combined to give $1/\tau_{\vec{q}j}^{ph-disloc} = 1/\tau_{\vec{q}j}^{ph-core} + 1/\tau_{\vec{q}j}^{ph-strain}$. We give values of the various parameters determined here for InN. $N_{ds}= N_d/42$, $N_{de}$=20 $N_d$/42, and $N_{dm}$=21 $N_d$/42 are defect densities determined from experiment for screw, edge and their mixture, respectively, as a function of the total defect density $N_d$ that varies for the different InN films. $b_s=c$ and $b_e = \sqrt{2}a/3$ are the Burger's vectors of screw and edge dislocations, respectively. $v_L$=5704 m/s is the calculated in-plane sound speed of the longitudinal phonon branch, $v_T$=2656 m/s is the sound speed of the transverse branch, $v_s$ is the average sound speed, $v_s^{-1} = (2/3v_T + 1/3v_L)^{-1}$. $\gamma$=2.132 is the calculated average Grüneisen parameter, $\nu$=0.330 is the Poisson ratio and $V = \sqrt{3}a^2c/8$ is the volume per atom. The Carruthers model [21]:

$$1/\tau_{\vec{q}j}^{ph-disloc} = \frac{1}{3} N_d b^2 \gamma^2 v_s |\vec{q}| [\log(1/b\sqrt{N_d})]^2 \quad (4)$$

given here after application of various approximations [21] has similar features to the Klemens model though depends explicitly on wavevector rather than frequency and is stronger due to the logarithmic term that represents a longer range strain field perturbation. Here, the Burger's vector $b$ is set to that of the edge dislocation, $b=b_e$.

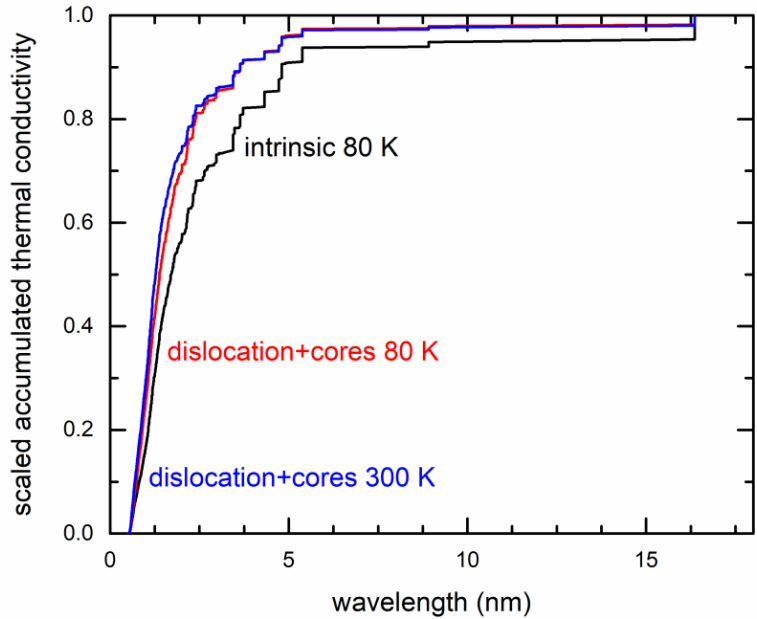

**Fig. S3-1 Accumulated thermal conductivity of intrinsic InN at 80 K (black), InN with dislocations and cores at 80 K (red) and 300 K (blue).**

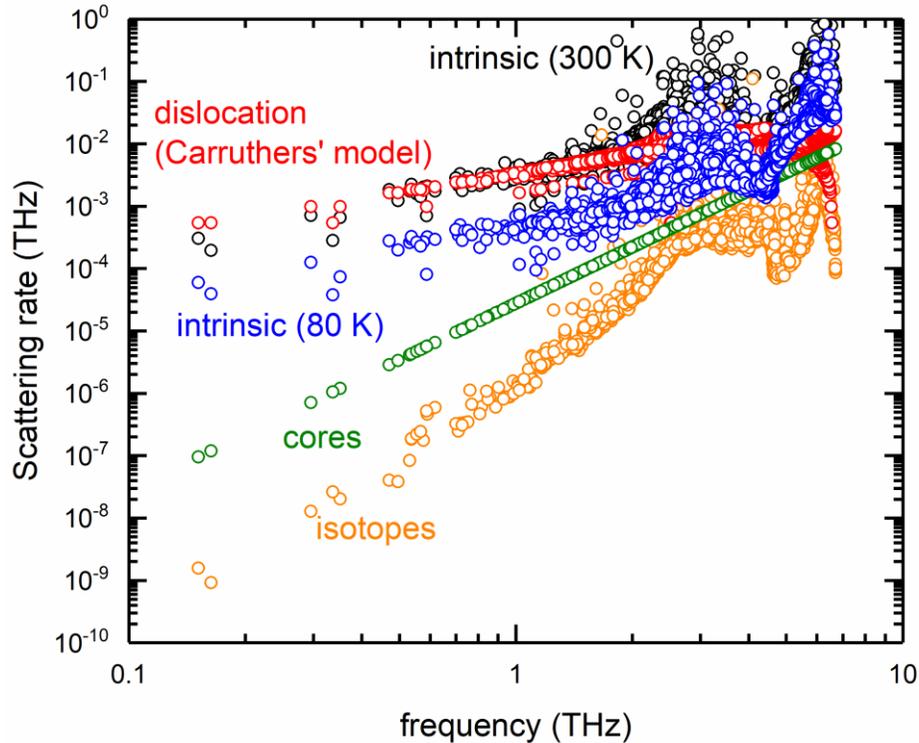

**Fig. S3-2 Scattering rates of dislocations using Carruthers' model (red), isotopes (orange) and dislocation cores (olive), in comparison with intrinsic three-phonon scattering in InN at 300 K (black) and 80 K (blue).**